\documentclass{aa501}
\usepackage{psfig}
\def\gtrsim{\lower.45ex\hbox{$\;\buildrel>\over\sim\;$}}
\def\ltrsim{\lower.45ex\hbox{$\;\buildrel<\over\sim\;$}}

\begin{document}


%
\title{The deeply embedded starburst in SBS~0335-052}
\thanks{Based on data obtained at ESO VLT UT1 on Cerro Paranal, Chile}

\author{ L. K. Hunt\inst{1}, L. Vanzi\inst{2}, T. X. Thuan\inst{3}}

\offprints{L.K. Hunt}

\institute{ CAISMI-CNR, Largo E. Fermi 5, 50125 Firenze - ITALY
email: hunt@arcetri.astro.it
\and
European Southern Observatory (ESO),
Alonso de Cordova 3107, Santiago - CHILE\\
email: lvanzi@eso.org
\and
Astronomy Department, University of Virginia, Charlottesville, 
VA 22903 - U.S.A.  email: txt@virginia.edu
}

\date{Received ; accepted }

\titlerunning{The deeply embedded starburst in SBS~0335-052}
\authorrunning{Hunt, Vanzi, \& Thuan}

\abstract{We present 4\,$\mu$m ISAAC imaging and spectroscopy
of the extremely metal--poor dwarf galaxy SBS~0335-052, aimed
at a better understanding of the dust in this low-metallicity galaxy.
The 4\,$\mu$m emission turns out to be very compact, 
confined to the brightest pair of Super Star Clusters (SSCs).
The {\it Ks--L$^\prime$} color is extremely red, and the 
$L^\prime$ emission
is consistent with the extrapolation of the ISO mid-infrared
spectral energy distribution (SED).
From hydrogen recombination lines and a fit to the near-/mid-infrared
SED, we confirm a visual extinction of $\gtrsim$ 15 mag.
Our data suggest that the sites of the optical and infrared emission
are distinct: the optical
spectral lines come from an almost dustless 
region with a high star formation rate and a few thousand
OB stars. 
This region lies 
along the line-of-sight to a very dusty central star cluster in which there
are more than three times as many massive stars, completely hidden
in the optical.
From the extinction, we derive an upper limit for the
dust mass of $10^{5}$ M$_\odot$ which could
be produced by recent supernovae.
}

\maketitle

%

\section{Introduction}

With an abundance of $Z_{\odot}/41$, 
SBS~0335-052 is the lowest-metallicity galaxy 
in the Second Byurakan Survey (Markarian et al. 1983),
and the second lowest known after I~Zw~18. 
Because of its low optical luminosity ($M_B=-16.7$),
its small size (3-4 kpc diameter), and its strong narrow
HII-region-like emission lines,
it is classified as a Blue Compact Dwarf (BCD) galaxy 
(Thuan et al. 1997).
SBS~0335-052 hosts an exceptionally powerful episode of star 
formation that, according to Thuan et al. (1997),
occurs mainly in six Super-Star Clusters (SSCs) not older than 25 Myr. 
ISO observations of SBS~0335-052 
(Thuan et al., 1999 - hereafter TSM) have revealed copious mid-infrared 
emission (the spectrum peaks at 14\,$\mu$m)
and a spectral energy distribution (SED) that 
is well fit from 7 to 17 $\mu$m by a heavily absorbed modified blackbody. 
From their fit TSM deduce an
optical extinction in the range 19-21 mag and speculate that most of the
star formation in SBS~0335-052 may be optically obscured. 


To better probe the extinction in SBS~0335-052,
Vanzi et al. (2000 - hereafter VHTI) obtained a high spatial 
resolution image in the {\it Ks} band and a near-infrared (NIR)
spectrum. 
Comparing these observations
with HST images and optical ground-based spectroscopy, 
they found no clear evidence for high extinction in SBS~0335-052.
However, NIR colors suggest that a fraction of the $K$-band flux
is produced by dust. 
Indeed that fraction is consistent with the
mid-infrared SED, and a fit of the combined from 2 to 15\,$\mu$m spectrum
gave a visual extinction $A_V\,\sim\,12$ magnitudes (VHTI).

To determine the nature of the dust, unexpected in such
a low-metallicity object, we have targeted SBS~0335-052 
at 4\,$\mu$m to obtain, for the first time,
spatially resolved imaging at a thermal wavelength.
At the same time, the low extinction at 4\,$\mu$m
enables us to probe the ionized gas via the 
Br$\alpha$ recombination line. 

\section{Observations}

\begin{figure}
\hspace{-0.5cm}\psfig{figure=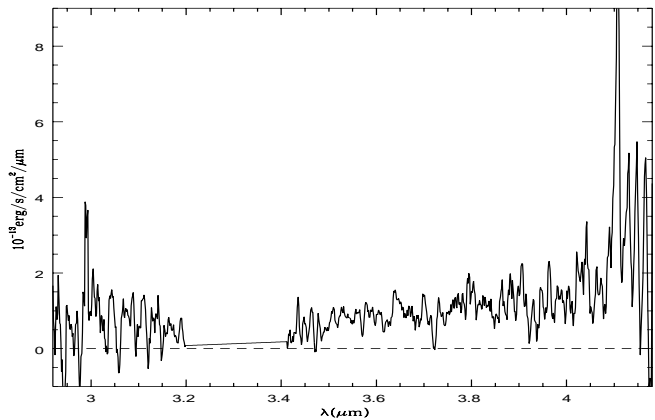,width=9.5cm,height=6cm}
\caption{ISAAC spectrum of SBS~0335-052. 
The horizontal dashed line shows the positions of zero flux, 
making evident the rise of the continuum toward long wavelengths.
The region of poor atmospheric transmission around 3.3\,$\mu$m
has been removed.
\label{spettro}
}
\end{figure}

We obtained a long-wavelength (2.7 - 4.2\,$\mu$m) low-resolution 
(R=360) spectrum of SBS~0335-052 with ISAAC at the ESO-ANTU (UT1). 
The observations were acquired in three different occasions on
October 9, 2000 (34 minutes integration) and on January 3 and 5, 2001 
(1 hour integration each). 
We used a 1\arcsec\ wide slit to match the already available NIR and
optical spectra and a position angle PA=145$^\circ$. 
The spectrum was acquired with an elementary integration
time of 0.104 sec, while chopping with the secondary 
and switching (North-South) between two beams with the telescope.
The telescope was dithered along the slit after each beam-switching cycle.
HR\,1891, a B2.5V star, was observed in the same way, in order to
eliminate the telluric absorption of the atmosphere. 
We used the Eclipse package developed by ESO to reduce the spectra,
following the standard procedures. 
The 2h34m combined spectrum, extracted with a 1\arcsec\ aperture,
is displayed in Fig. \ref{spettro}. 
Since the dispersion is 14\AA/pix, but the slit is almost 7 pixels wide, 
the original 
spectrum has been smoothed to the effective resolution dictated by the slit.
The spectrum was flux-calibrated with the photometry 
of our $L$-band image described below.

We used the same instrument and telescope to also acquire an image 
in the $L$\footnote{Because of its central wavelength $\neq$ 3.5\,$\mu$m,
it is more appropriate to call this band $L^\prime$,
which we will do hereafter.}
band at 3.8\,$\mu$m ($\Delta\lambda~=~0.58\,\mu$m).
The pixel scale was 0.0709\arcsec/pixel and the 
total integration time 30 minutes, with an  
on-chip integration of 0.104 sec. 
Like the spectrum, the image was acquired  
while chopping with the secondary 
and nodding the telescope between the two beams. 
The telescope was also dithered randomly after each beam-switching cycle.
We used the IRAF package for the data reduction, and 
relied on the telescope offsets to align the dithered images.
The seeing FWHM measured from a star in the {\bf negative} image is 
6.9 pixels, or 0.49\arcsec .
The $L^\prime$ image is shown in Fig. \ref{image}.

The $L^\prime$-band photometry was calibrated with the photometric
standard HD\,22686, 
obtained in the same way as the target, and
assuming an $L^\prime$ magnitude of 7.20 (Elias et al. 1982).
The growth curve of the object in the $L^\prime$ image is shown in
the upper panel of Fig. \ref{growth}; the total $L^\prime$ magnitude is 
14.1\,$\pm$\,0.2, shown as a horizontal dotted line.

\subsection{Photometry and colors}

\begin{figure}
\centerline{\psfig{figure=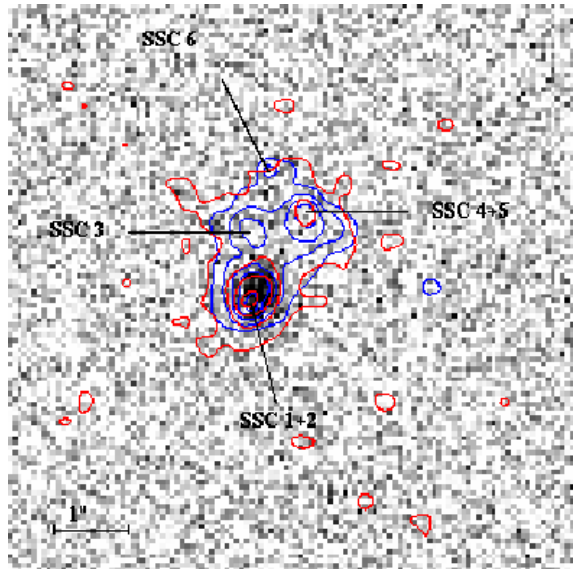,width=7.5cm,clip=}}
\caption{{\bf $L^\prime$}-band ISAAC/VLT image 
with superimposed contours of the {\it Ks} image in red
(VHTI), 
and those of the HST $V$ image in blue (Thuan et al. 1997)
(N up, E left).
The SSCs denoted by Thuan et al. (1997) are also labeled.
}
\label{image}
\end{figure}

SBS~0335-052 consists, as noted in the Introduction, of several
SSCs superimposed on an extended blue underlying gaseous envelope.
The two brightest SSC groups, SSC~1$+$2 and SSC~4$+$5
(using the notation of Thuan et al. 1997),
are separated by approximately 1.4\,\arcsec.
The registration of the $L^\prime$ image is
made difficult by the lack of field stars in the field-of-view
of the direct image.
Nevertheless, because of the N-S nodding direction, a relatively
bright star $\sim$\,26\arcsec\ E and 52\arcsec\ S appears in the
image of the negative beam.
We have therefore used this star in the Digital Sky Survey (DSS)
image to register the $L^\prime$ frame, since the star is too
far away to appear in either the HST or the {\it Ks} image (VHTI).
The astrometry relative to this star indicates that the $L^\prime$
source corresponds to the brightest SSCs, SSC~1$+$2, although
with $\sim$\,0.5--0.7\,\arcsec\ uncertainty because of the large
pixels (1.7\,\arcsec ) of the DSS.
Superimposed on our $L^\prime$ image in
Fig. \ref{image} are the contours of the HST image (blue) 
and the high-resolution {\it Ks} image (red). 

\begin{figure}
\psfig{figure=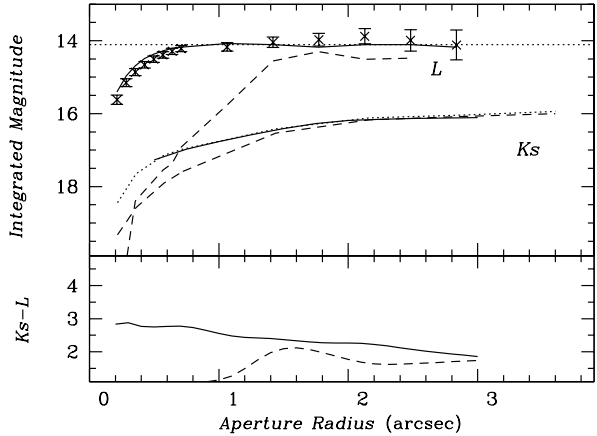,width=8.8cm,clip=}
\caption{$L^\prime$-band growth curve and {\it Ks--$L^\prime$} 
cumulative profile of SBS~0335-052.
The top panel shows the data points, with respective errors, and the
horizontal dotted line the total magnitude of 14.11;
a solid line illustrates the growth
curve for a point source on the final image.
Also shown in the top panel is the {\it Ks} growth curve derived from
the high-resolution {\it Ks} image in VHTI.
The bottom panel shows the {\it Ks--$L^\prime$} color. 
For all curves, the SSC~1$+$2 component is shown as a
solid line, and SSC~4$+$5 as a dashed one.
}
\label{growth}
\end{figure}

To verify the association of the $L^\prime$ source with SSC~1$+$2,
we have plotted in Fig. \ref{growth}
the $L^\prime$ and {\it Ks} growth curves centered on SSC~1$+$2 and SSC~4$+$5, 
the pairs of SSCs visible in the shorter-wavelength images.
Because there was no $L^\prime$ emission at the SSC~4$+$5 peak, 
we derived the $L^\prime$ photometry by fixing the position to
the {\it Ks} SSC~4$+$5 peak. 
It is clear from Fig. \ref{growth}
that the radial trend of the $L^\prime$ emission is consistent
with the {\it Ks} SSC~1$+$2 position, since the $L^\prime$ emission 
from the SSC~4$+$5 is initially linear with radius, which is what is 
expected for no signal at the nominal center.
(The SSC~4$+$5 curve then turns over at roughly the separation
of the two SSC pairs.) 

Figure \ref{growth} (lower panel)
shows the {\it Ks--$L^\prime$} color obtained from combining
our $L^\prime$ photometry with that from the {\it Ks} image in VHTI.
The {\it Ks--$L^\prime$} color is extremely red, ranging from 2.8 at the center
to roughly 2 for the total\footnote{Because these curves are cumulative,
such a trend implies a radial blueward gradient; the 2\,$\mu$m emission
is more extended than that in $L^\prime$.}.
After correcting the color for our redder $L^\prime$ filter 
(see Bessell \& Brett 1988), we find $K-L$ colors of 2.1 to about 1.5
(these colors correspond to an $L$ filter centered at 3.5\,$\mu$m).
Such a red color is highly unusual in (non-Seyfert)
extra-galactic objects, and is
redder than all but one of the HII galaxies studied by Glass \& Moorwood (1985).
Indeed, the exception, NGC~5253, has $K-L$ colors similar to SBS~0335-052,
and has been called the ``youngest starburst known'' by Rieke et al. (1988).
Because it is also a low-metallicity dwarf galaxy, we will use NGC~5253
below as a ``benchmark'' for further comparisons.
These colors have not been corrected for the ionized gas emission
(see Section \ref{emission}).

\section{The dust in SBS~0335-052}

The dust reponsible for the $L^\prime$ emission appears to
be {\it very compact}, confined to a region at 
most 1.2\arcsec\ in diameter (see Fig. \ref{growth}).
At a distance of 54.3~Mpc (Thuan et al. 1997), this corresponds
to approximately 300~pc.
(This estimate is consistent with the physical extent of the 
ISO--emitting dust hypothesized by TSM.)
The FWHM of the 4\,$\mu$m emission is roughly
equivalent to the seeing FWHM, namely 0.5\arcsec\
or 130~pc.
As seen in Fig. \ref{growth},
the $L^\prime$ curve is very slightly more extended than a point source,
but is much more compact than the {\it Ks} flux, which reaches
its asymptotic value at a diameter of roughly 1~kpc (4\arcsec).
(SBS~0335-052 is actually more extended than this at very low
$K$ surface brightness levels, see VHTI).

\subsection{Extinction  \label{ext}}

Brackett $\alpha$ in emission is clearly detected in our spectrum:
we measure a flux of $9.0\pm1~10^{-15}~$erg/s/cm$^2$ in an aperture of 
1\,\arcsec$\times$1.5\,\arcsec . 
We can compare this value with the fluxes measured for
Br$\gamma$ by VHTI and H$\beta$ by Izotov et al. (1997) in the same
aperture, and use the ratio to derive the visual extinction
(assuming a foreground screen). 
Using the intrinsic line ratios given in Osterbrock (1989) appropriate
for SBS~0335-052 (Case B, 20000~K, 400-500/cm$^3$: Izotov et al. 1997),
together with the extinction curve from Cardelli et al. (1989), 
we obtain from the H$\beta$/Br$\alpha$ ratio 
$A_V=1.45\pm0.11$\,mag. 
The Br$\gamma$/Br$\alpha$ ratio gives  $A_V=12.1\pm1.8$. 
Izotov et al (1997) measured $A_V=0.55$\,mag from H$\alpha$/H$\beta$, 
and Vanzi et al. (2000) $A_V=0.73$ from H$\beta$/Br$\gamma$. 
There is therefore a
clear tendency of the extinction to increase with wavelength. 

\begin{figure}
\psfig{figure=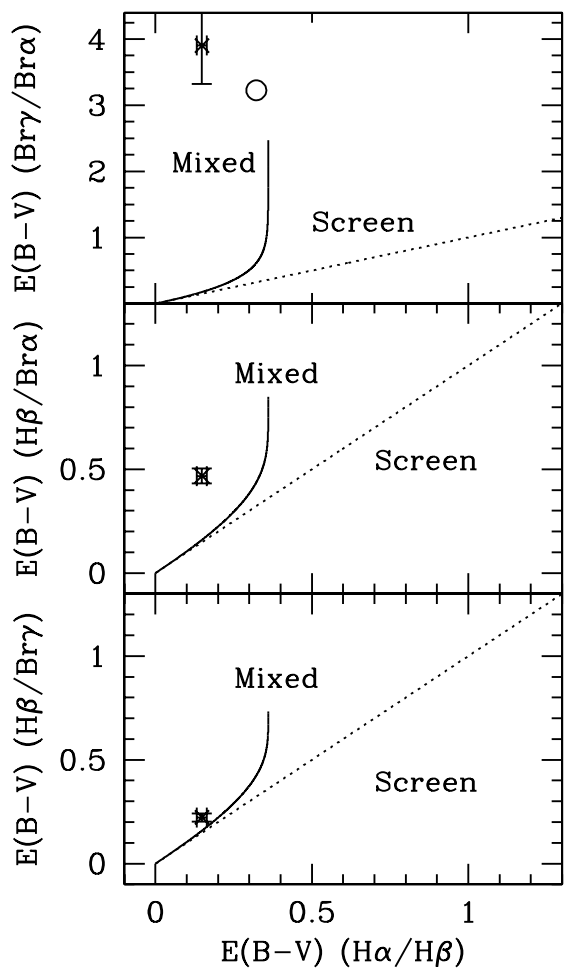,width=8cm}
\caption{The color excess $E(B-V)$ derived from the
hydrogen recombination line ratios; 
the data from SBS\,0335-052 are shown as $\times$.
Two simple models are also illustrated:
the foreground screen model (labeled ``Screen'') and a homogeneous mixture of
gas and dust (labeled ``Mixed'').
The open circle shows data for NGC~5253 (Beck et al. 1996).
\label{tau}
}
\end{figure}

With different assumptions,
recombination line ratios can be used to infer either the interstellar
extinction curve, or the geometry of the obscuring dust.
In the latter case, longer wavelength line ratios have the virtue of
probing deeper into embedded regions, an important advantage for
dusty starbursts.
Following Calzetti et al. (1996), we have calculated
the color excess $E(B-V)$ derived from the hydrogen
recombination lines.
As before,
we adopted the Cardelli et al. (1989) expression for the 
interstellar extinction curve, and the intrinsic line
ratios from Osterbrock (1989).
The color excesses are shown in Fig. \ref{tau},
together with two dust models, 
a foreground dust screen and a homogenously mixed slab of gas and dust.
It is clear from the figure that a foreground screen model is 
highly inappropriate both for SBS\,0335-052 ($\times$) and 
for NGC~5253 (open circle, taken from Beck et al. 1996).
It is also evident that the model of homogeneous mixed slab 
predicts an increase of $A_V$ with wavelength, qualitatively similar 
to our observations.
Nevertheless, while 
NGC~5253 appears to be well described by homogeneously mixed gas and dust, 
the low H$\alpha$/H$\beta$ ratio in SBS~0335-052 together 
with the extremely high infrared line ratios are not well
reproduced by a single value of $\tau$ in the 
mixed-medium model.\footnote{A clumped medium (Natta \& Panagia 1984)
is even worse at reproducing the observed line ratios.} 

Because the extinction derived from the optical line ratios
is small ($A_V\,=\,0.55$\,mag), a more probable model consists of
a nearly dustless region, responsible for virtually all of the
optical emission, lying in front of a highly obscured central
knot.
We exclude the possibility that
the interstellar extinction curve could be grossly incorrect,
since we have also used the Seaton (1979) and Landini et al. (1984)
curves and obtained similar results.
An extinction law with a higher $R=A_V/E(B-V)$
coupled with a mixed medium model could fit the data.
Such ratios are typical of denser environments such as dark
molecular clouds (e.g., Kim et al. 1994), but the applicability 
of such a law to SBS~0335-052 is not clear.
Therefore, in Section 4 we will characterize
the starburst in SBS~0335-052
using the simple two-component model outlined above, 
namely a low-$A_V$ region in front of a region highly obscured.

\subsection{Emission \label{emission}}

We have added new 4\,$\mu$m and 2\,$\mu$m points
(VHTI) to the ISO spectrum of SBS~0335-052 (TSM),
and fit the resulting SED.
First, though, we had to estimate what fraction of the $K$ and 
$L^\prime$ flux is due to dust. 
It turns out that the gas fractions determined in VHTI and below, 
together with reasonable estimates of stellar colors, 
constrain the relative contributions of ionized gas, dust, and stars 
from 1.6 to 4\,$\mu$m.
The integrated {\it Ks} magnitude was used by VHTI 
to compare with the ISO observations obtained with a beam size
larger than the galaxy; here, 
because of the compact 4\,$\mu$m morphology,
we adopt the VHTI $Ks$ photometry for SSC~1$+$2 only. 
With the emission coefficients given in Joy \& Lester (1988),
and our Br$\alpha$ observation, we find that 27\% of the $L^\prime$
emission derives from the ionized gas continuum. 
Then assuming: 
{\it i)}~no dust emission in $H$ (VHTI);
{\it ii)}~stellar $H-K\,\sim\,0.0-0.2$;
{\it iii)}~stellar $K-L^\prime\,\sim\,0.0-0.5$,
we derive a stellar fraction of
$\,\sim\,$37\% at 2\,$\mu$m, 
compared with $\,\sim\,$6\% at 4\,$\mu$m\footnote{The quoted
values are the mean of the color range cited above.}.
For simplicity, we have not considered any 
obscuration of the stellar component.
With a 50\% gas fraction at 2\,$\mu$m (VHTI),
we therefore obtain a dust-emitting fraction of 13\% at 2\,$\mu$m,
and 67\% at 4\,$\mu$m,
corresponding to 0.014~mJy at $K$ and 0.37~mJy at $L^\prime$.

\begin{figure}
\psfig{figure=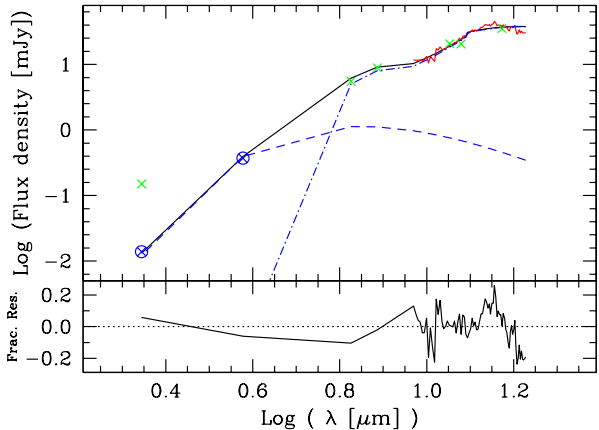,width=8.8cm,clip=}
\vspace{-2cm}
\caption{Near- and mid-infrared SED of SBS~0335-052
with residuals from the best fit.
In the upper panel, the fit is shown by a solid line, while
the dashed line shows the warm modified blackbody (449~K)
which is not absorbed.
The dot-dashed line shows the cool modified blackbody (193~K),
and affected by a foreground screen of $A_V\,=\,$16.3 mag.
The 2\,$\mu$m point lying above the fit ($\times$) 
illustrates the $K$ value used in VHTI, roughly 10 times higher
than that adopted here.
The lower panel illustrates the fractional residuals which are
largest around 12-14\,$\mu$m, exactly where the extinction curve
is most uncertain.
}
\label{dustfit}
\end{figure}

The combined SED from 2 to 17\,$\mu$m was fit with two modified
blackbodies (MBBs); the cooler one is obscured by 
dust in a foreground screen with an extinction curve given by 
Lutz (1999). 
As shown by TSM, this curve provides a much better
fit than previous infrared extinction curves, because of its
significantly higher extinction in the 3--8\,$\mu$m region.
The emissivity was fixed to $\lambda^{-1.5}$ (TSM).
Figure \ref{dustfit} shows the resulting fit;
the warmer (unobscured, shown by a dashed line) 
MBB has a temperature of 459$\,\pm\,20$~K,
and the cooler one (dot-dashed line) 192$\,\pm\,4$~K, obscured
by dust with $A_V\,=\,16.3\,\pm\,0.5$~mag. 
These results differ slightly from VHTI, but should be more reliable
because of the additional constraint of the 4\,$\mu$m data point,
and the use of a more refined extinction curve; 
also, in VHTI the emissivity was left as a free parameter,
and the 2\,$\mu$m flux was higher because it included emission
from the whole galaxy not just SSC~1$+$2 (see above).
In any case, the fit is very similar to that found by TSM
and confirms the high extinction found by them.

Although our data suggest that the geometry is probably more a 
homogeneously mixed
medium than a foreground screen (see Fig. \ref{tau}), 
the comparable values of the extinction 
($A_\lambda/A_V\,=\,0.06-0.08$\,mag)
in the MIR spectral region (Lutz 1999)
make the shapes of the two fits very similar, 
but with $A_V$ 4 to 5 times larger in the mixed medium model.
We therefore adopted the foreground screen model since it gave
fits of similar, if not better, quality than the mixed medium model. 

It was not possible to fit the SED with two MBBs
absorbed (``extincted'') by the same amount, because the resulting 
fit falls off too steeply toward short wavelengths.
Since the temperature and amplitude of the warm MBB
are basically governed by only the 2 and 4\,$\mu$m data points,
an additional quantity such as extinction would not be constrained. 
However, the data do constrain the extinction 
for the warm MBB to be much smaller than that for the cool one
(assuming the extinction curve is correct).
Such a result is also consistent with the two-component model of the
previous section.

\subsection{Mass \label{mass}}

The mid-infrared spectrum of SBS~0335-052 is unusual compared to other 
star-forming galaxies (e.g., Dale et al. 2001),
because of its positive 3--4\,$\mu$m slope, its strong continuum,
and lack of spectral features (e.g., Unidentified Infrared 
Bands: UIBs; Polycyclic Aromatic Hydrocarbons: PAHs). 
Its dust properties are therefore not well described by standard
dust models (D\'esert et al. 1990), especially regarding the
very-small-grain component, and as a result we cannot plausibly estimate 
the mass of the {\it emitting} dust.
Nevertheless, the extinction is likely governed by the cooler dust,
and we can estimate the mass of the obscuring dust from the $A_V$ derived
from our fit of the SED. 

With $A_V\,=\,16.3$~mag, 
and assuming N(HI)/$A_V\,\sim\,2\times10^{21}\,{\rm cm}^2$/mag
and a gas-to-dust mass ratio of $\sim$\,200,
we derive a dust surface density of 1.2~$M_\odot$/pc$^2$.
This calculation assumes solar metallicity, but the
metallicity correction factors cancel out in the end.
The derived dust surface density is similar to that of the 
model of Draine \& Lee (1984) which, with a grain opacity
of 3000 cm$^2$/gm, is 1.6~$M_\odot$/pc$^2$.
If the extinction arises in a volume similar in size
to the emission region (diameter 1.2\arcsec), 
we obtain a dust mass of $\sim\,10^5$\,$M_\odot$.
Alternatively, we can use the formalism of Spitzer (1978) to 
relate the amount of matter required to produce the observed 
extinction to the mean extinction curve, the density of solid 
material within the grain, and the grain dielectric constant 
(Aannestad \& Purcell 1973).
From the observed extinction, and assuming the grain
parameters of Draine \& Lee (1984),
we can therefore derive the mean volume density of the dust 
$\rho_d\,=\,0.0065\,M_\odot$/pc$^3$, which is more than 
a factor of 20 higher than the Galactic value of 
$0.0003 M_\odot$/$pc^3$.
In a spherical region of diameter 1.2\arcsec\ as above,
the dust mass becomes $\sim\,10^5$\,$M_\odot$,
consistent with the previous estimate.
These estimates are both upper limits since we have assumed a
uniform distribution of the dust within the 1.2\arcsec\ region;
the actual value may be less if the dust does not extend over
the entire region.
These upper limits are comparable with those obtained by TSM.


\section{Properties of the starburst}

Our measurement of the Br$\alpha$ line shows that a considerable 
fraction of the star formation in SBS~0335-052 is optically obscured, 
invisible even at 2\,$\mu$m. 
Assuming the two-component geometry described in Section \ref{ext},
we can adopt the (extinction corrected)
H$\beta$ flux measured by Izotov et al. (1997)
to estimate the Br$\gamma$ and Br$\alpha$ flux
originating in the same (low $A_V$) region.
This exercise yields values that are respectively about 50 and 25\% of 
what is observed. 
That is to say, 50\% of the observed Br$\gamma$ flux and
25\% of the Br$\alpha$ comes from the (virtually)
dustless foreground region which is the origin of the
H$\beta$ emission.
The remaining
50\% of the Br$\gamma$ flux and 75\% of Br$\alpha$
arise in the embedded dusty star cluster;
in the optical
we are only observing about 1/4 of the total ionized gas emission. 

The fraction of optically hidden emission can be used to estimate 
the extinction in the embedded cluster.
We obtain $A_V\,\sim\,$15, close to that inferred
from the fit of the mid-infrared SED.
The bright Brackett $\alpha$ line, corrected for $A_V\,\sim\,$15,
indicates that the total star formation rate in the embedded 
cluster, instead of the optically-derived value of 
0.4\,$M_\odot$/yr (Thuan et al. 1997), is more 
like $\sim$\,1.7\,$M_\odot$/yr.
This last value is more than 40 times higher than that in the 
lowest-metallicity BCD I~Zw~18 (0.04\,$M_\odot$/yr).

The number of massive stars in the embedded star cluster
can be inferred from the the Br$\alpha$ luminosity.
With 75\% of the observed $Br\alpha$ flux coming from
the obscured region, and correcting as above for $A_V\,\sim\,$15,
$L_{Br\alpha}$(embedded)~=~$4.2\times10^{39}$ erg/s.
Then, using the
prescription of Guseva et al. (2000), we obtain
approximately $\sim$ 14200 O7 stars. 
There are {\bf more than three times} as many massive stars in the
embedded cluster as in the unobscured region
($\sim 4040$, with the observed $L_{H\beta}$ luminosity
of $1.9\times10^{40}$ erg/s in the same aperture: VHTI).
The total mass of the embedded cluster can be estimated
by assuming an Initial Mass Function (IMF);
adopting the IMF given by Scalo (1998) gives a mass of
1.2$\times 10^7\,M_\odot$ in the embedded cluster.
Alternatively, with a Salpeter IMF, and lower and upper
mass cutoffs of respectively 0.8 and 120\,$M_\odot$ (Schaerer \& Vacca 1998),
we obtain an embedded stellar mass of 6.6$\times 10^6\,M_\odot$.
The stellar density in the embedded cluster
turns out to be $\sim$\,85--150\,$M_\odot$/pc$^2$, which is 
relatively low compared to the SSCs in NGC~5253
with a stellar surface density of $\gtrsim 10^3\,M_\odot$/pc$^2$
(Calzetti et al. 1997).

The massive star cluster in the obscured central knot
would be expected to host a significant number of Type II supernovae (SNe).
Our Br$\alpha$ measurement, together with the lowest-metallicity
models of Leitherer et al. (1999), implies a SN rate of 0.004-0.006/yr,
according to the stellar mass adopted (see above).
We can therefore
calculate the number of SNe expected to reside in the region of the SSCs,
which turns out to be $\sim\,$8000-12000, when integrated over the
lifetime of the burst after the onset of SNe.
If each metal-poor SN produces, on average, 1 $M_\odot$ of dust
(Todini \& Ferrara 2001), we would expect a dust mass on the order
of $10^4 M_\odot$, a factor of 10 lower than
the upper limit derived from the extinction.
This may suggest that the dust is not spread out uniformly over the
1.2\arcsec\ region.
Also, the dust mass we infer from the extinction is
highly uncertain because of the extremely low metallicity of SBS~0335-052.
The physical conditions in the dense dusty medium of
the embedded star cluster do not resemble those of the
solar neighborhood, nor are they similar to the cold
dark clouds in the Galaxy.
No dust models exist for extremely low-metallicity environments;
even the Magellanic Clouds modelled by Weingartner \& Draine
(2001) have metallicities about 4 times higher than that in
SBS~0335-052.
The grain properties such as the size 
distribution or the chemical composition may be radically different
in low-metallicity environments.
It is also true that even in solar-metallicity contexts
the canonical models are still a subject of debate.
There is evidence, for example, that a more realistic size
distribution may produce the same extinction
with a lower dust mass (Kim et al. 1994).
Also, some fraction of grains may be ``fluffy'' 
which would also increase the extinction per unit dust mass
(Kr\"ugel \& Siebenmorgen 1994; Wolff et al. 1994; Mathis 1996;
Snow \& Witt 1996).

SBS~0335-052, while unusual for such a metal-poor object,
does have peers with similar properties, although at a higher metal abundance.
NGC~5253 ($\sim$1/5 $Z_\odot$) and Henize~2-10 ($\sim$1/10 $Z_\odot$)
are sub-solar metallicity low-luminosity dwarf galaxies
(the former a dwarf elliptical, the latter a BCD) 
which host powerful central starbursts with SSCs
(Calzetti et al. 1997; Conti \& Vacca 1994),
similar to SBS~0335-052.
Both galaxies show high extinction from the Brackett recombination
line ratios (Beck et al. 1996,  Kawara et al. 1989),
but low extinction from the optical lines 
(see Fig. \ref{tau}); they also show
significant 10\,$\mu$m absorption features (Kawara et al. 1989), and
host significant numbers of Wolf-Rayet 
stars in the central clusters (e.g., Guseva et al. 2000). 
Because of this and other evidence, the bursts are thought to be 
only a few Myr old
(Beck et al. 1996; Calzetti et al. 1997; Beck et al. 1997; VHTI).
In all three galaxies, the extinction is so high that
optical measurements cannot be used to reliably study the embedded
star clusters; even at 2\,$\mu$m a significant fraction of the ionized
gas emission is completely obscured.
In the higher-metallicity objects (NGC~5253 and He~2-10), it appears 
that the stellar 
clusters are born embedded in dust and then emerge after 2--3~Myr,
becoming bright in the optical and the UV (Calzetti et al. 1997).
Strictly speaking, such a scenario may not be appropriate
for SBS~0335-052, since, if the present burst is the first one, 
dust could not have predated the current episode of star formation.
The quantity of dust inferred from our measurements is roughly
compatible with that from primordial SN models, but the clusters 
themselves would have ``polluted'' the starburst. 
Either way, judging from SBS~0335-052 and its extremely
low metal abundance, it is probable
that much of primordial star formation could have occurred 
in deeply embedded dense star clusters.

\section{Conclusions}

\begin{enumerate}
\item We find direct evidence for a heavily absorbed central knot of
star formation in SBS~0335-052. 
The extinction measured from the the observed Br$\gamma$/Br$\alpha$ 
is $A_V\,\approx\,12$ mag,
but line-ratio comparisons indicate that there is a virtually dust-free
region along the line-of-sight to
massive star clusters which are enshrouded in dust.
\item The infrared SED has been fitted with two modified blackbodies,
and the inferred extinction of 16.3 mag is roughly consistent with,
but slightly larger than, that obtained from Br$\gamma$/Br$\alpha$.
The total amount of obscuring dust is estimated to be around 
 $10^5\,M_\odot$.
\item The star formation rate and number of massive stars
in the embedded cluster in SBS~0335-052 are more than three 
times higher than inferred from optical measurements.
Roughly 3/4 of the star
formation in SBS~0335-052 occurs within a highly obscured embedded cluster.
\item It is likely that the SNe produced by the current burst are
responsible for the dust present in SBS~0335-052, although rough estimates
of their dust production are lower than the upper limits 
for the dust mass inferred from extinction. 
\end{enumerate}

\begin{acknowledgements}
We are especially grateful to the ESO Director General, C. Cesarsky,
for the generous allocation of Director's Discretionary Time, as well
as the VLT-ISAAC staff who conducted the observations.
We would like to thank Marc Sauvage who passed on the digitial 
version of the ISO spectrum, and Dieter Lutz who gave us instantly
his digital version of the Galactic extinction curve. 
We also acknowledge an anonymous referee for a careful reading of
the manuscript and cogent comments which greatly improved the paper.
TXT thanks the partial financial support of NASA-JPL Contract 961535.
\end{acknowledgements}


\begin{thebibliography}{}

\bibitem[1973]{aane} Aannestad P.A., Purcell E.M. 1973, ARA\&A, 309

\bibitem[1996]{beck} Beck S.C., Turner J.L., Ho P.T.P., Lacy J.H., Kelly D.M.
1996, ApJ 457, 610

\bibitem[1997]{beck2} Beck S.C., Kelly D.M., Lacy J.H. 1997, AJ 114, 585

\bibitem[1988]{bessell}Bessell M.S., Brett J.M. PASP 100, 1134

\bibitem[1978]{bohlin}Bohlin R.C., Savage B.D., Drake J.F. 1978, ApJ
224, 132

\bibitem[1996]{calzetti} Calzetti D., Kinney A.L., Storchi-Bergmann T. 
1996, ApJ 458, 132

\bibitem[1997]{calzetti2} Calzetti D., Meurer G.R., Bohlin R.C. et al.
1997, AJ 114, 1834

\bibitem[1989]{cardelli} Cardelli J.A., Clayton G.C., Mathis J.S. 1989,
ApJ 345, 245

\bibitem[1994]{conti} Conti P.S., Vacca W.D. 1994, ApJ 423, 97

\bibitem[2001]{dale} Dale D.A., Helou G., Contursi A., Silbermann N.A., 
Kolhatkar S. 2001, ApJ 549, 215

\bibitem[1990]{desert}D\'esert F.-X., Boulanger F., Puget J.L. 1990, 
A\&A 237, 215

\bibitem[1984]{draine}Draine B.T., Lee H.M. 1984, ApJ 285, 89

\bibitem[1982]{elias}Elias J.H., Frogel J.A., Mathews K., Neugebauer G. 1982,
AJ 87, 1029

\bibitem[1985]{glass}Glass I.S., Moorwood A.F.M. 1985, MNRAS 214, 429

\bibitem[2000]{guseva}Guseva N.G., Izotov Y.I., Thuan T.X. 2000, ApJ 531, 776

\bibitem[1999]{hirashita}Hirashita H. 1999, ApJ 522, 220


\bibitem[1997]{Izotov97} Izotov Y. I., Lipovetsky V. A., Chaffee F. H. et al. 
1997, ApJ 476, 698

\bibitem[1988]{Joy} Joy M. \& Lester D. F. 1988, ApJ 331, 1451


\bibitem[1994]{kim}Kim S.-H., Martin P.G., \& Hendry P.D. 1994, ApJ 422, 164

\bibitem[1994]{krugel}Kr\"ugel E. \& Siebenmorgen R. 1994, A\&A 288, 929

\bibitem[1984]{landini} Landini M., Natta A., Oliva E., Salinari P.,
Moorwood A.F.M. 1984, A\&A 134, 284

\bibitem[1999]{sb99} Leitherer C., Schaerer D., Goldader J. D., et al. 1999, ApJS 123, 3


\bibitem[1996]{lutz} Lutz D. 1999, The Universe as Seen by ISO, Eds.
P. Cox \& M.F. Kessler, ESA-SP 427, 623

\bibitem[1983]{sbs} Markarian B.E., Lipovetsky V.A., Stepanian J.A. 1983,
Astrofizika 19, 29

\bibitem[1996]{mathis}Mathis J.S. 1996, ApJ 472, 643

\bibitem[1989]{osterbrock} Osterbrock D.E. 1989, {\it Astrophysics of Gaseous
Nebulae and Active Galactic Nuclei}, University Science Books, Mill Valley


\bibitem[1988]{rieke88} Rieke G.H., Lebofsky M.J., Walker C.E. 1988, ApJ 325, 679

\bibitem[1998]{scalo}Scalo J. 1998, ASP Conf. Series 142, 201

\bibitem[1998]{schaerer}Schaerer D. \& Vacca W.D. 1998, ApJ 497, 618

\bibitem[1979]{seaton} Seaton J.J. 1979, MNRAS 187, 73P

\bibitem[1996]{snow}Snow T.P. \& Witt A. 1996, ApJ 468, 65

\bibitem[1978]{spitzer}Spitzer L. Jr. 1978, {\it Physical Processes in
the Insterstellar Medium}, John Wiley \& Sons, Inc., New York

\bibitem[1997]{Thuan97} Thuan T. X., Izotov Y. I., Lipovetsky V. A.
1997, ApJ 477, 661

\bibitem[1999]{Thuan99} Thuan T. X., Sauvage M., Madden S. 1999,
ApJ 516, 783 (TSM)

\bibitem[2001]{todini} Todini P., Ferrara A. 2001, MNRAS, 325, 726

\bibitem[2000]{Vanzi00} Vanzi L., Hunt L. K., Thuan T. X., Izotov Y. I. 2000, 
A\&A 363, 493 (VHTI)

\bibitem[2001]{weingartner}Weingartner J.C. \& Draine B.T. 2001, ApJ 548, 296

\bibitem[1994]{wolff}Wolff M.J., Clayton G.C., Martin P.G., \& Schulte-Ladbeck
R.E. 1994, ApJ 423 412

\end{thebibliography}
\end{document}